\documentclass[twocolumn,showkeys,aps,prb,showpacs]{revtex4-1}
\usepackage{graphicx}
\usepackage[CJKbookmarks,dvipdfm,colorlinks,linkcolor=blue,citecolor=blue]{hyperref}

\begin{document}

\title{Biaxial  strain enhanced  piezoelectric properties in monolayer  g-$\mathrm{C_3N_4}$}

\author{San-Dong Guo$^{1,2}$, Wen-Qi Mu$^{1}$ and Yu-Tong Zhu$^{1}$}
\affiliation{$^1$School of Electronic Engineering, Xi'an University of Posts and Telecommunications, Xi'an 710121, China}
\affiliation{$^2$Key Laboratary of Advanced Semiconductor Devices and Materials, Xi'an University of Posts and Telecommunications, Xi'an 710121, China }
\begin{abstract}
Graphite-like carbon nitride (g-$\mathrm{C_3N_4}$) is considered  as a promising candidate
for energy materials. In this work, the biaxial  strain (-4\%-4\%) effects  on  piezoelectric properties  of g-$\mathrm{C_3N_4}$ monolayer   are studied by density functional theory (DFT). It is found that  the increasing strain can reduce the elastic coefficient  $C_{11}$-$C_{12}$, and increases  piezoelectric stress  coefficient  $e_{11}$, which lead to the enhanced  piezoelectric  strain  coefficient $d_{11}$. Compared to unstrained one, strain of  4\% can  raise  the $d_{11}$  by about 330\%. From -4\% to 4\%, strain can  induce  the improved ionic contribution  to $e_{11}$ of g-$\mathrm{C_3N_4}$,  and almost unchanged electronic contribution,  which is different from $\mathrm{MoS_2}$ monolayer (the enhanced electronic contribution and reduced  ionic contribution).  To prohibit  current leakage, a piezoelectric material  should  be a semiconductor, and g-$\mathrm{C_3N_4}$ monolayer is always a semiconductor in considered strain range. Calculated results show that the gap  increases from compressive strain to tensile one. At 4\% strain, the first and second valence bands cross, which has important effect on transition dipole moment (TDM).
Our works provide  a strategy to achieve enhanced  piezoelectric effect of g-$\mathrm{C_3N_4}$ monolayer, which gives a useful guidence for developing efficient energy conversion devices.

\end{abstract}
\keywords{g-$\mathrm{C_3N_4}$, Piezoelectronics, 2D materials}

\pacs{71.20.-b, 77.65.-j, 72.15.Jf, 78.67.-n ~~~~~~~~~~~~~~~~~~~~~~~~~~~~~~~~~~~Email:sandongyuwang@163.com}

\maketitle

\section{Introduction}
Analogous to graphene,  monolayer  g-$\mathrm{C_3N_4}$ has been achieved  by top-down methods, which  provides several  potential
applications including  superior photocatalytic activities\cite{c2,c2-1},
sensing\cite{c3} and memory devices\cite{c4}. For example, g-$\mathrm{C_3N_4}$ can generate
hydrogen from water under visible light with an appropriate band gap  of 2.7
eV\cite{c5}. Many g-$\mathrm{C_3N_4}$/semiconductor heterostructures have been constructed   to improve the
photocatalytic performance of g-$\mathrm{C_3N_4}$ to restrain the
recombination of photogenerated carriers, like  g-$\mathrm{C_3N_4}$/$\mathrm{MoS_2}$\cite{c6},  g-$\mathrm{C_3N_4}$/$\mathrm{TiO_2}$\cite{c7} and g-$\mathrm{C_3N_4}$/CdS\cite{c8}.  Because of  non-centrosymmetric structure, the monolayer  g-$\mathrm{C_3N_4}$  can  exhibit a piezoelectricity, which may produce  potential piezocatalysis applications.

In fact, due to  potential nanoscale piezoelectric
applications, the piezoelectricities of two-dimensional (2D) materials have
attracted growing interest\cite{q4}.
 Experimentally,   the piezoelectric  coefficient ($e_{11}$=2.9$\times$$10^{-10}$ C/m) of the monolayer $\mathrm{MoS_2}$ has been measured
  with  the 2H phase\cite{q5,q6}, and an intrinsic vertical piezoelectric response\cite{q8} has been proved to exist in the Janus MoSSe monolayer.
The theoretical studies on piezoelectric properties of  2D materials, such as transition metal dichalchogenides (TMD), Janus TMD, group IIA and IIB metal oxides, group-V binary semiconductors and group III-V semiconductors,  have been widely stuided \cite{q7,q7-1,q7-2,q7-3,q7-4,q9,q10,q11,q12,qr}.
The giant piezoelectricities  in monolayer SnSe,
SnS, GeSe and GeS  have been reported,  as high as  75-251 pm/V\cite{q10}.   A only in-plane piezoelectricity exits in many 2D materials, for example TMD monolayers\cite{q9}, and an additional out-of-plane piezoelectricity has also been predicted in many  2D  Janus materials\cite{q7,q7-2,q7-3}.
A  pure  out-of-plane piezoelectric response has been predicted in  penta-graphene\cite{q7-4}, and two strategies are proposed to enhance its piezoelectric properties by strain and constructing Janus monolayer.
 The strain effects on   the  piezoelectric response of  $\mathrm{MoS_2}$\cite{r1}, AsP\cite{q7-1}, SnSe\cite{q7-1} and Janus TMD monolayers\cite{r3} have been reported, and their piezoelectric properties can be effectively tuned. For example,   the $d_{22}$  of SnSe monolayer at -3.5\% strain along the armchair  direction is up to 628.8 pm/V  from unstrained 175.3 pm/V\cite{q7-1}.

\begin{figure}
  \includegraphics[width=7.0cm]{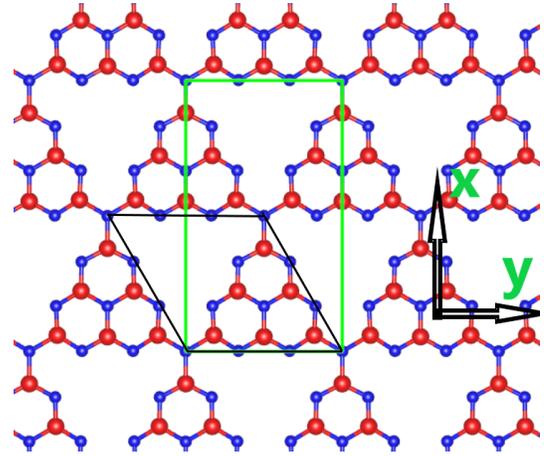}
  \caption{(Color online)The crystal structure of monolayer  g-$\mathrm{C_3N_4}$, and the primitive cell is
   are marked by black line. The large red balls represent C atoms, and the  small blue balls for N atoms. The rectangle supercell
   is marked by green line to calculate piezoelectric coefficients.  }\label{t0}
\end{figure}

In this work,  the biaxial  strain-tuned piezoelectric properties of g-$\mathrm{C_3N_4}$ monolayer  are studied by using density functional perturbation theory (DFPT)\cite{pv6}.  Only  in-plane piezoelectricity   exists for  g-$\mathrm{C_3N_4}$ monolayer. The independent piezoelectric constants $d_{11}$ is predicted to be 1.42 pm/V. It is found that tensile strain of 4\% can improve $d_{11}$ to 6.12 pm/V, which is due to reduced $C_{11}$-$C_{12}$ and enhanced $e_{11}$. Different from $\mathrm{MoS_2}$, increasing strain can enhance the  ionic contribution  to $e_{11}$. It is found that the gap of g-$\mathrm{C_3N_4}$ increases from -4\% to 4\% strain. Strain can also induce the cross between the first and second valence bands, producing important effects on TDM. Therefore, our works give an experimental proposal  to achieve enhanced  piezoelectricity in g-$\mathrm{C_3N_4}$ monolayer.

The rest of the paper is organized as follows. In the next
section, we shall give our computational details and methods  about piezoelectric coefficients.
 In the third section, we perform symmetry analysis for elastic and piezoelectric coefficients. In the fourth  sections, we shall present main results of g-$\mathrm{C_3N_4}$ monolayer. Finally, we shall give our  conclusions in the fifth section.

\begin{table}
\centering \caption{For monolayer  g-$\mathrm{C_3N_4}$, the lattice constants $a_0$ ($\mathrm{{\AA}}$), the elastic constants $C_{ij}$ ($\mathrm{Nm^{-1}}$), shear modulus
$G_{2D}$ ($\mathrm{Nm^{-1}}$),  Young's modulus $C_{2D}$  ($\mathrm{Nm^{-1}}$),  Poisson's ratio $\nu$,  the HSE06 gaps  (eV) and piezoelectric coefficients   $e_{11}(sum)$ [the  electronic  $e_{11}(ele)$ and ionic $e_{11}(ion)$ contribution] ($10^{-10}$ C/m ) and  $d_{11}$ (pm/V),  with previous theoretical values and experimental results
given in parentheses and square brackets. }\label{tab0}
  \begin{tabular*}{0.48\textwidth}{@{\extracolsep{\fill}}cccc}
  \hline\hline
$a_0$& $C_{11}$ &  $C_{12}$& $G_{2D}$\\\hline
7.134 (7.135\cite{m1})[7.130\cite{c5}] &184.92&48.02&68.45\\\hline\hline
$C_{2D}$& $\nu$& $Gap$& $e_{11}(sum)$\\\hline
172.45&0.26&2.77(2.76\cite{m1})[2.7\cite{c5}] & 1.94 (2.18\cite{nc})\\\hline\hline
$e_{11}(ele)$& $e_{11}(ion)$& $d_{11}$&\\\hline
3.02&-1.08&1.42 & \\\hline\hline

\end{tabular*}
\end{table}

\begin{figure*}
  \includegraphics[width=13cm]{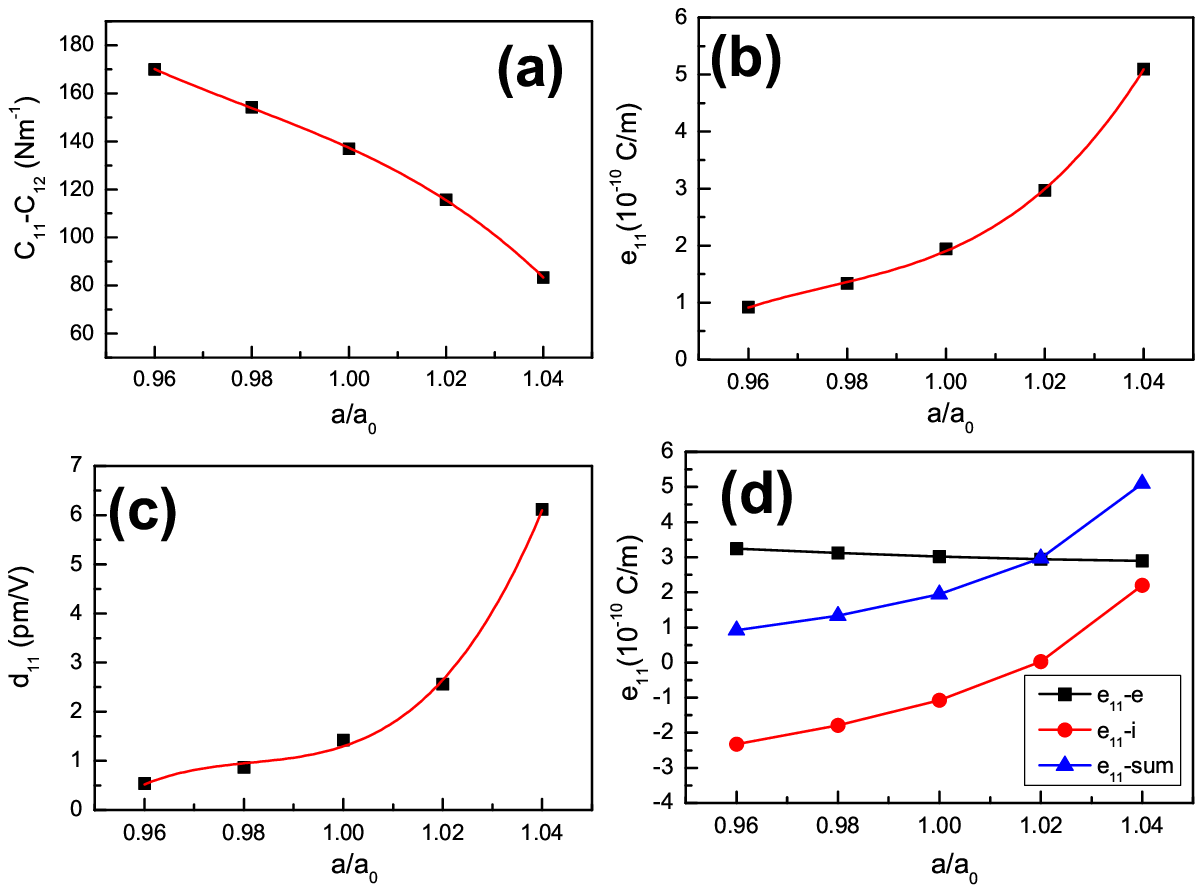}
  \caption{(Color online) For monolayer g-$\mathrm{C_3N_4}$, the elastic constants (a) $C_{11}$-$C_{12}$,  piezoelectric coefficients (b) $e_{11}$ and (c) $d_{11}$, and (d) the ionic contribution and electronic contribution to $e_{11}$  as a function of  biaxial  strain.}\label{cn}
\end{figure*}

\section{Computational detail}
 Within the framework of DFT\cite{1}, we carry out our calculations  by using the  VASP code\cite{pv1,pv2,pv3} with the projected augmented wave
(PAW) method. A kinetic cutoff energy of 500 eV is adopted, and  we use the popular GGA of Perdew, Burke and  Ernzerhof  (GGA-PBE)\cite{pbe} as the exchange-correlation potential to calculate piezoelectric and elastic properties. To avoid interactions
between two neighboring images, a vacuum spacing of
 more than 19 $\mathrm{{\AA}}$ along the z direction is added.
The total energy  convergence criterion is set
to $10^{-8}$ eV, and  the Hellmann-Feynman forces  on each atom are less than 0.0001 $\mathrm{eV.{\AA}^{-1}}$.
The  elastic stiffness tensor  $C_{ij}$  and  the piezoelectric stress coefficients $e_{ij}$ are calculated by using strain-stress relationship (SSR) and   DFPT method\cite{pv6}.
 Within SSR and DFPT, the electronic and ionic contribution to
the elastic and  piezoelectric stress coefficients can be calculated directly from VASP code.
 The Brillouin zone sampling of  g-$\mathrm{C_3N_4}$ monolayer
is done using a Monkhorst-Pack mesh of 11$\times$11$\times$1  for $C_{ij}$, and  6$\times$11$\times$1 for $e_{ij}$.
The Heyd-Scuseria-Ernzerhof (HSE06) hybrid functional with default parameters
is used to obtain the correct electronic structures of monolayer  g-$\mathrm{C_3N_4}$. The TDM are calculated by using VASPKIT code\cite{vk}.
The 2D elastic coefficients $C^{2D}_{ij}$
 and   piezoelectric stress coefficients $e^{2D}_{ij}$
have been renormalized by the the length of unit cell along z direction ($Lz$):  $C^{2D}_{ij}$=$Lz$$C^{3D}_{ij}$ and $e^{2D}_{ij}$=$Lz$$e^{3D}_{ij}$.

\section{Symmetry Analysis}
In noncentrosymmetric crystals,   a change of polarization can be induced  by strain or stress. The phenomenon can be described by the third-rank piezoelectric stress tensors  $e_{ijk}$ and strain tensor $d_{ijk}$, which are from the sum of ionic
and electronic contributions:
 \begin{equation}\label{pe0}
      e_{ijk}=\frac{\partial P_i}{\partial \varepsilon_{jk}}=e_{ijk}^{elc}+e_{ijk}^{ion}
 \end{equation}
and
 \begin{equation}\label{pe0-1}
   d_{ijk}=\frac{\partial P_i}{\partial \sigma_{jk}}=d_{ijk}^{elc}+d_{ijk}^{ion}
 \end{equation}
In which  $P_i$, $\varepsilon_{jk}$ and $\sigma_{jk}$ are polarization vector, strain and stress, respectively.
For 2D materials, if we only consider in-plane strain components\cite{q7,q9,q10,q11,q12} using Voigt notation,
 the  $d_{ij}$ can be derived  by  the  relation:
 \begin{equation}\label{pe}
  \left(
    \begin{array}{ccc}
      e_{11} & e_{12} & e_{16} \\
     e_{21} & e_{22} & e_{26} \\
      e_{31} & e_{32} & e_{36} \\
    \end{array}
  \right)
  =
  \left(
    \begin{array}{ccc}
      d_{11} & d_{12} & d_{16} \\
      d_{21} & d_{22} & d_{26} \\
      d_{31} & d_{32} & d_{36} \\
    \end{array}
  \right)
    \left(
    \begin{array}{ccc}
      C_{11} & C_{12} & C_{16} \\
     C_{21} & C_{22} &C_{26} \\
      C_{61} & C_{62} & C_{66} \\
    \end{array}
  \right)
   \end{equation}
The elastic tensor  $C_{ij}$  can be calculated by  SSR,  and the  $e_{ij}$  can be attained by DFPT.
The space group number of monolayer g-$\mathrm{C_3N_4}$  is 187, and the corresponding point group $\bar{6}m2$ reduces   $e_{ij}$,  $d_{ij}$ and $C_{ij}$ into:
 \begin{equation}\label{pe1}
  \left(
    \begin{array}{ccc}
      e_{11} &-e_{11} & 0 \\
    0 &0 & -e_{11}\\
      0 & 0 & 0 \\
    \end{array}
  \right)
  \end{equation}
  \begin{equation}\label{pe1}
  \left(
    \begin{array}{ccc}
        d_{11} & -d_{11} & 0 \\
    0 &0 & -2d_{11} \\
      0 & 0 & 0 \\
    \end{array}
  \right)
  \end{equation}
  \begin{equation}\label{pe1}
    \left(
    \begin{array}{ccc}
      C_{11} & C_{12} &0 \\
     C_{12} & C_{11} &0 \\
     0 & 0 & \frac{C_{11}-C_{12}}{2} \\
    \end{array}
  \right)
   \end{equation}
Here, the only in-plane $d_{11}$ is derived by  \autoref{pe}:
\begin{equation}\label{pe2-7}
    d_{11}=\frac{e_{11}}{C_{11}-C_{12}}
\end{equation}

\section{Main calculated results}
The structure of the
monolayer g-$\mathrm{C_3N_4}$ is illustrated  in \autoref{t0}, which shows a large aperture and a triazine ring as a unit.
Firstly,  the optimized lattice constants of monolayer g-$\mathrm{C_3N_4}$ is $a$=$b$=7.134 $\mathrm{\AA}$ using GGA, which agrees well with previous theoretical and experimental values\cite{m1,c5}. The band structure of
the g-$\mathrm{C_3N_4}$ monolayer using HSE06 is calculated,
which shows an indirect band gap semiconductor with  the valence band maximum (VBM) at $\Gamma$ point and
the conduction band minimum (CBM)  at K point. The position of CBM is different from previous one (at M point)\cite{m1}, which may be due to different
HSE06 parameters.
The calculated HSE06 band gap is 2.77 eV, which is very close to
 the experimental value (2.7 eV)\cite{c5} and the previous
calculated result (2.76 eV)\cite{m1}.
The independent elastic stiffness coefficients of $C_{11}$ and $C_{12}$ are calculated, and the
monolayer has  constants of $C_{11}$=184.92 $\mathrm{Nm^{-1}}$  and  $C_{12}$=48.02 $\mathrm{Nm^{-1}}$, which meet the  Born  criteria of mechanical stability.  These elastic constants are larger than ones of $\mathrm{MoS_2}$\cite{q9,q11}.  The 2D Young¡¯s moduli $C^{2D}$   and shear modulus $G^{2D}$ can be expressed as\cite{ela}:
\begin{equation}\label{e1}
C^{2D}=\frac{C_{11}^2-C_{12}^2}{C_{11}}
\end{equation}
\begin{equation}\label{e1}
G^{2D}=C_{66}
\end{equation}
\begin{equation}\label{e1}
C_{66}=\frac{C_{11}-C_{12}}{2}
\end{equation}
The corresponding Poisson's ratios is given:
\begin{equation}\label{e1}
\nu^{2D}=\frac{C_{12}}{C_{11}}
\end{equation}
The calculated values are $C^{2D}$=172.45 $\mathrm{Nm^{-1}}$, $G^{2D}$=68.45 $\mathrm{Nm^{-1}}$  and $\nu^{2D}$=0.26. The  related data  of g-$\mathrm{C_3N_4}$  monolayer are listed in \autoref{tab0}.

\begin{figure*}
  \includegraphics[width=13cm]{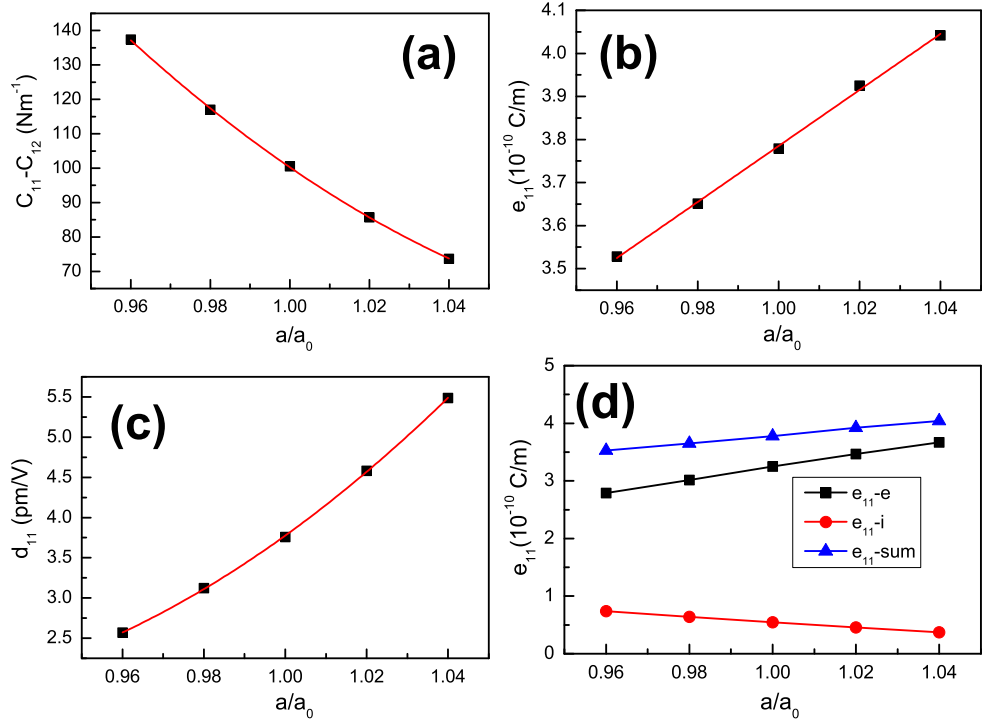}
  \caption{(Color online)For monolayer $\mathrm{MoS_2}$, the elastic constants (a) $C_{11}$-$C_{12}$,  piezoelectric coefficients (b) $e_{11}$ and (c) $d_{11}$, and (d) the ionic contribution and electronic contribution to $e_{11}$  as a function of  biaxial  strain.}\label{cn1}
\end{figure*}

The rectangle supercell is used to calculate piezoelectric stress coefficients   of g-$\mathrm{C_3N_4}$  monolayer by DFPT, and the x and y directions are shown in \autoref{t0}. The calculated piezoelectric coefficient $e_{11}$=1.94$\times$$10^{-10}$ C/m, being close to previous value $e_{11}$=2.18$\times$$10^{-10}$ C/m\cite{nc}. It is found that   the  electronic contribution  is opposite to ionic contribution, and they are 3.02 $\times$$10^{-10}$ C/m and -1.08 $\times$$10^{-10}$ C/m, respectively.
Based on calculated $e_{11}$, $C_{11}$ and $C_{12}$, the predicted  $d_{11}$ is 1.42 pm/V, which is smaller than  most 2D  TMD monolayers\cite{q11}.
Strain  strategy  is an effective method   to improve  piezoelectric effect  of 2D materials\cite{q7-1,r1,r3}. Here, we only consider biaxial  strain, which can not produce polarization, not like uniaxial strain. In the simulation, the small  biaxial   strain (-4\% to 4\%) effects on  piezoelectric properties of monolayer g-$\mathrm{C_3N_4}$ are studied, which may be easily achieved in experiment. The elastic constants  $C_{11}$-$C_{12}$,  piezoelectric coefficients  $e_{11}$ and  $d_{11}$ as a function of  biaxial  strain are plotted in \autoref{cn}.
It is clearly seen that the  $C_{11}$-$C_{12}$ decreases, and $e_{11}$ increases, when the strain changes from -4\% to 4\%. This will lead to improved
$d_{11}$  according to \autoref{pe2-7} with 4\% to 4\% strain.
 At 4\% strain, the $d_{11}$ becomes  6.12 pm/V   from unstrained 1.42 pm/V, increased  by 331\%.

The ionic contribution and electronic contribution to $e_{11}$  as a function of  biaxial  strain are also plotted in \autoref{cn}.
It is found that there are  narrow variations for electronic contribution with -4\% to 4\% strain, and only varies -0.35$\times$$10^{-10}$ C/m.
However, the magnitude change  of the ionic contribution is very large, and about 4.52$\times$$10^{-10}$ C/m. Therefore, the ionic contribution has an important role to enhance piezoelectric effect  of g-$\mathrm{C_3N_4}$  monolayer caused by strain, which is different from a typical 2D piezoelectric material $\mathrm{MoS_2}$. The $\mathrm{MoS_2}$ monolayer has the same point group $\bar{6}m2$ with g-$\mathrm{C_3N_4}$, which gives rise to the same reduced piezoelectric coefficients. The elastic constants  $C_{11}$-$C_{12}$,  piezoelectric coefficients  $e_{11}$ and  $d_{11}$, and the ionic contribution and electronic contribution to $e_{11}$ of monolayer $\mathrm{MoS_2}$ as a function of  biaxial  strain are plotted in \autoref{cn1}.
 For unstrained $\mathrm{MoS_2}$, our calculated $C_{11}$ (131.76 $\mathrm{Nm^{-1}}$), $C_{12}$ (31.20 $\mathrm{Nm^{-1}}$), $e_{11}$ (3.78$\times$$10^{-10}$ C/m) and $d_{11}$ (3.76 pm/V) agree well with previous theoretical values (130 $\mathrm{Nm^{-1}}$, 32 $\mathrm{Nm^{-1}}$, 3.64$\times$$10^{-10}$ C/m, 3.73 pm/V)\cite{q11}.
 For  $C_{11}$-$C_{12}$,   $e_{11}$ and  $d_{11}$, the change trend is similar to one of g-$\mathrm{C_3N_4}$ with strain changing from -4\% to 4\%. However, the  electronic part has positive  contribution to improve piezoelectric effect  of  $\mathrm{MoS_2}$,   while the ionic part gives negative effect. Thus,  the electronic part dominate the enhancement of piezoelectric effect  of monolayer $\mathrm{MoS_2}$ caused by strain.

\begin{figure*}
   \includegraphics[width=16cm]{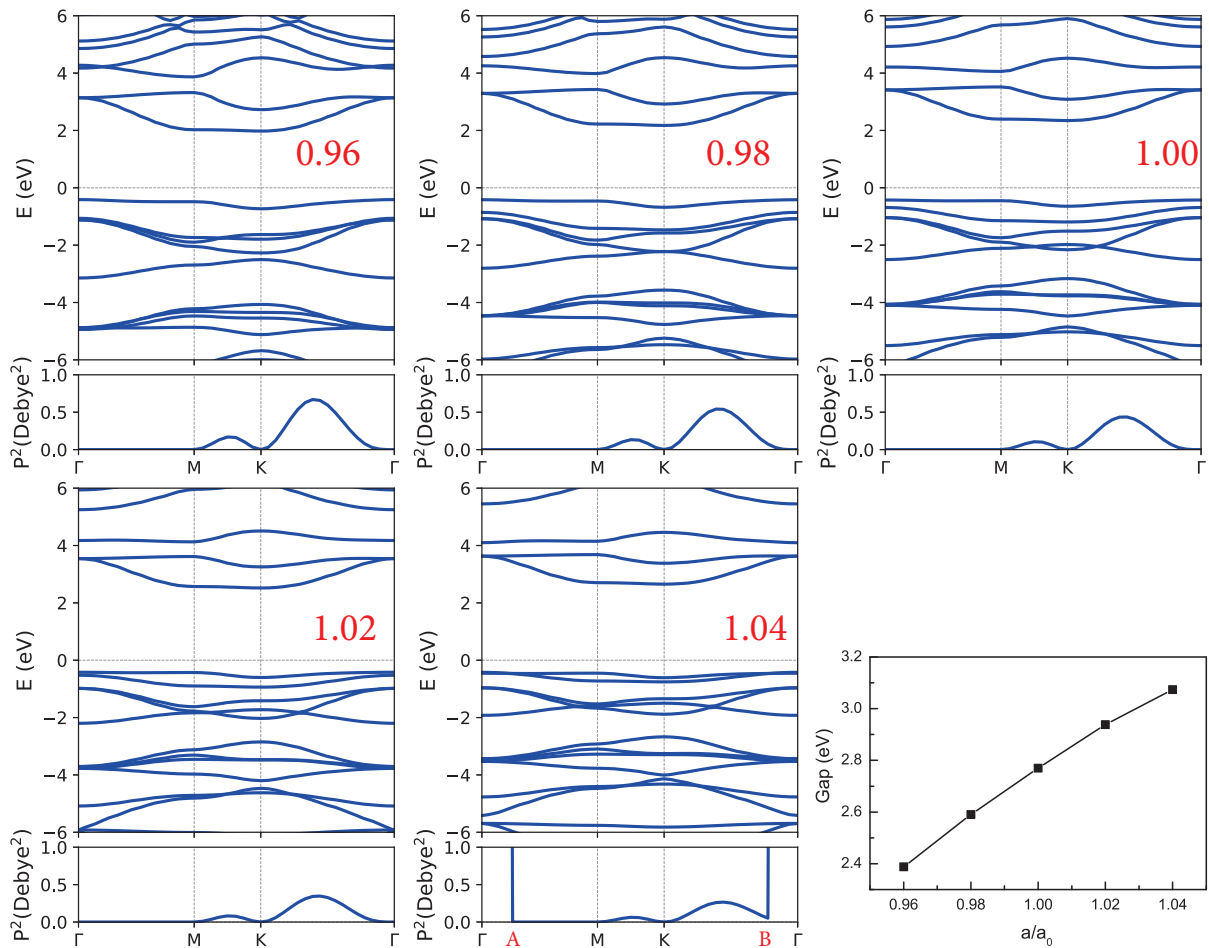}
  \caption{(Color online)The energy band structures, TDM and energy band gap (Gap) of   monolayer g-$\mathrm{C_3N_4}$  using HSE06  with the application of  biaxial strain (-4\%-4\%).}\label{neng}
\end{figure*}

The monolayer g-$\mathrm{C_3N_4}$ at applied strain,  exhibiting  piezoelectricity,  not only  should   break inversion symmetry,  but also should have a band gap. To confirm strained g-$\mathrm{C_3N_4}$ to be a semiconductor, the energy band structures and  gaps using HSE06 as a function of strain are plotted in \autoref{neng}.  It is clearly seen that the gap increases from 2.39 eV (-4\%) to 3.07 eV (4\%), but the positions of VBM and CBM do not change.
It is found that the strain can induce the cross between the first and second valence bands at about 4\% strain, which produces important effect on TDM.
The TDM  is the electric dipole moment associated with the transition between the two states, and  we calculate the squares of TDM from the highest valence band  to the lowest conduction band, which is also shown in \autoref{neng}. The calculated results show that the outline of TDM has little change from -4\% to 2\%, which are mainly along  $\Gamma$-K  and K-M.  However, the magnitude of TDM  becomes huge along  $\Gamma$-A and  $\Gamma$-B at 4\% strain because of band cross between the first and second valence bands. So, strain can also produce important influence on  optical absorptions of monolayer g-$\mathrm{C_3N_4}$.

\section{Conclusion}
In summary, the  reliable first-principles calculations are performed to investigate the biaxial  strain (-4\%-4\%) effects on piezoelectric properties in monolayer  g-$\mathrm{C_3N_4}$.  Compared to unstrain one, compressive strain reduces $e_{11}$, and increases $C_{11}$-$C_{12}$. However, tensile strain produces opposite effects on $e_{11}$ and $C_{11}$-$C_{12}$. These lead to improved $d_{11}$ from compressive strain to tensile one. Calculated results show that the ionic contribution  to $e_{11}$ of g-$\mathrm{C_3N_4}$ is in favour of the strain-induced enhanced $d_{11}$, which is different from  $\mathrm{MoS_2}$ monolayer. It is found that the HSE06 gap increases from 2.39 eV (-4\%) to 3.07 ev (4\%). The tensile strain (4\%) can induce the cross between the first and second valence bands, which can induce huge TDM.
Our predictive findings can  provide a simple  way to achieve energy-efficient energy transformation devices.

\begin{acknowledgments}
This work is supported by the Natural Science Foundation of Shaanxi Provincial Department of Education (19JK0809). We are grateful to the Advanced Analysis and Computation Center of China University of Mining and Technology (CUMT) for the award of CPU hours and WIEN2k/VASP software to accomplish this work.
\end{acknowledgments}

\end{document}